\documentclass{pasa}%

\def\lsim{\mathrel{\rlap{\lower 3pt \hbox{$\sim$}} \raise 2.0pt \hbox{$<$}}}
\def\gsim{\mathrel{\rlap{\lower 3pt \hbox{$\sim$}} \raise 2.0pt \hbox{$>$}}}

\title[Growth of the First Black Holes]{The Early Growth of the First
  Black Holes}
\author[Johnson \& Haardt]{Jarrett L. Johnson$^1$\thanks{jlj@lanl.gov}
  and Francesco Haardt$^2$ \\
\affil{$^1$X Theoretical Design, Los Alamos National Laboratory, Los
  Alamos, NM 87545, USA}%
\affil{$^2$DiSAT, Universit\'a dell'Insubria, via Valleggio 11, 22100 Como, Italy}}%
\jid{PASA}
\doi{10.1017/pas.\the\year.xxx}
\jyear{\the\year}

\begin{document}%
\begin{abstract}
With detections of quasars powered by increasingly massive black holes
(BHs) at increasingly early times in cosmic history over the past
decade, there has been correspondingly rapid progress made on the theory of early
BH formation and growth.  Here we review the emerging picture of how
the first massive BHs formed from the primordial gas and then grew to 
supermassive scales.  We discuss the initial conditions for the
formation of the progenitors of these seed BHs, the factors dictating
the initial masses with which they form, and their initial stages
of growth via accretion, which may occur at super-Eddington rates.
Finally, we briefly discuss how these results connect to large-scale
simulations of the growth of supermassive BHs over the course of the first billion
years following the Big Bang.
\end{abstract}
\maketitle%
\section{INTRODUCTION}
\label{sec:intro}
Supermassive black holes (SMBHs) reside in the centers of massive galaxies
(e.g. Kormendy \& Ho 2013), and the presence of quasars at the highest
redshifts (e.g. Willott et al. 2003; Fan et al. 2006; Mortlock et al. 2011; Venemans
et al. 2013) powered by some of the most massive BHs (e.g. Wu et
al. 2015) suggests that this has been the case since the dawn of
galaxy formation.  It is a subject of intense
research how these BHs initially formed and subsequently grew so
rapidly (see previous reviews by e.g. Volonteri 2010, 2012; Volonteri \& Bellovary 2012; 
and Alexander \& Hickox 2012; Haiman 2013).     

The formation of the first massive BHs begins with the
collapse of the primordial gas to form their stellar progenitors.
Depending on the environment in which these progenitors form, the mass
scale of the seed BHs they leave behind can vary wildly, from $\sim$
10 to $\sim$ 10$^5$ M$_{\odot}$.  In turn, the rate at which these BHs
initially grow is determined by their initial masses and the conditions
of the gas reservoirs from which they accrete.  While most BHs which
form in the early universe are expected to grow at relatively modest
rates, a small fraction may grow extremely rapidly, either due
to their large initial masses and or to accretion flows which
are heavy enough to trap the radiation emitted in the accretion process.    
The final fates of the first massive BHs are dictated by the
availability of gas to accrete on cosmological scales and by the
impact of the merging and coalescence of these BHs with one another.

Here we review recent theoretical exploration of these processes, from 
the formation of the first massive seed BHs, to the early stages of
their growth and how they meet their fates as the supermassive BHs 
powering high redshift quasars.   

\begin{figure*}
\begin{center}
\includegraphics[width=3.6in,angle=-90]{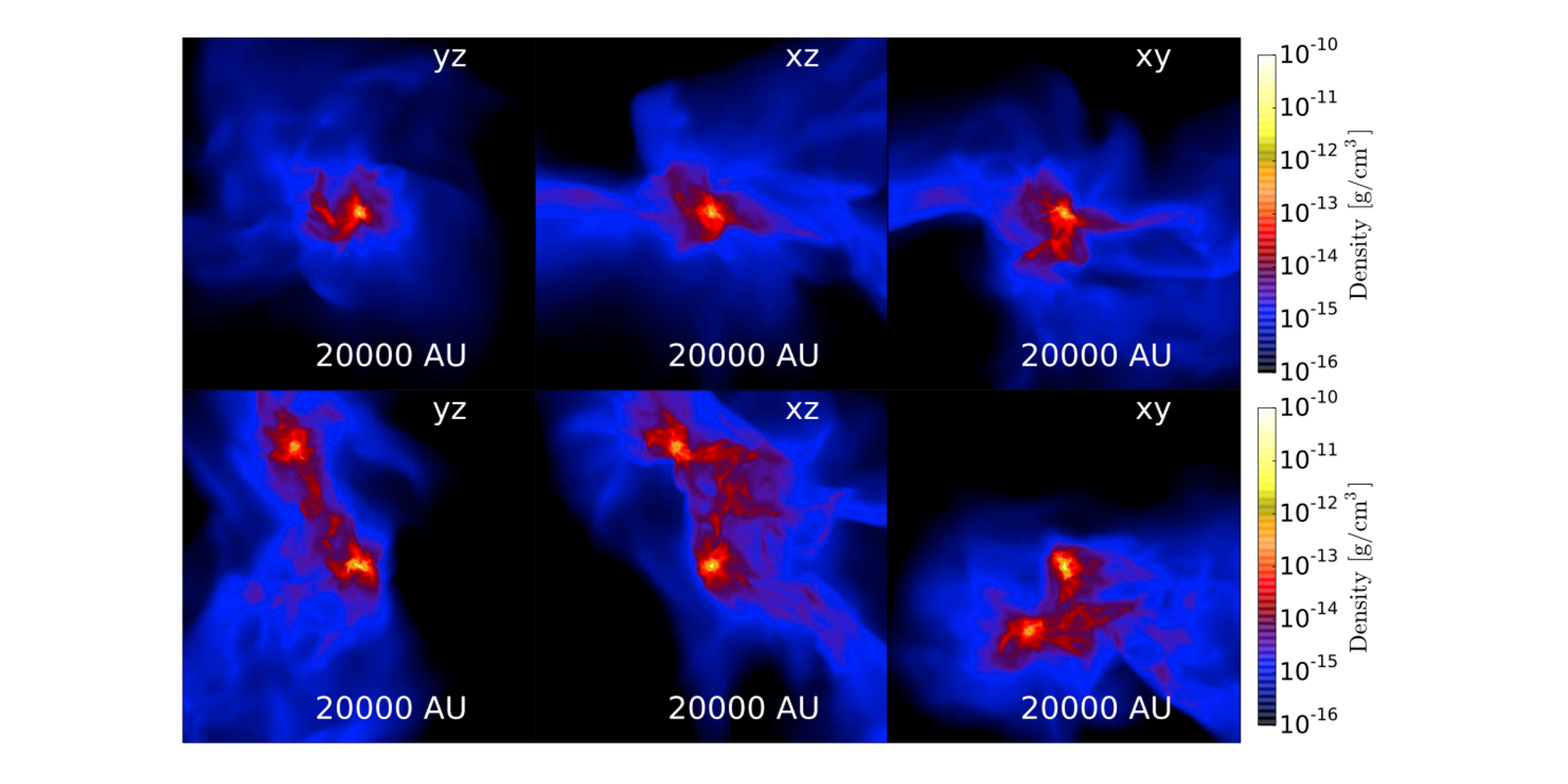}
\caption{The projected gas density in halos hosting one (top row) and two (bottom row) rapidly growing primordial protostars, along the three axes in a cosmological simulation, as labeled.  At the early stages of accretion shown here, the objects have masses a few times that of the Sun, but they are expected to grow rapidly to become $\ge$ 10$^4$ M$_{\odot}$ supermassive stars within a few million years.  From Latif et al. (2015b).
}\label{Fig3}
\end{center}
\end{figure*}

\section{Initial Formation of the First Massive Black Holes} 
We begin by discussing the factors determining the masses to which primordial
stars grow prior to their collapse into the first massive BHs.  We
consider first the impact of the environment on the growth of the
first stars.  We then discuss the processes which determine their final masses.

\subsection{Impact of External Sources of Radiation}
The first stars form from the collapse of the primordial gas, a
process which is facilitated by the radiative cooling of
both atomic and molecular hydrogen (H$_{\rm 2}$) (e.g. Bromm \& Larson
2004; Grief 2015).  Which of these two species of hydrogen dominates the
cooling rate largely dictates the mode of star formation and the
masses of the stars which form.  

If molecular hydrogen forms in sufficient abundance, the primordial 
gas is expected to cool to $\sim$ 100 K and to collapse into
protostars which accrete at rates of 10$^{-4}$ - 10$^{-3}$
M$_{\odot}$ yr$^{-1}$ (e.g. Yoshida et al. 2008).  At these accretion
rates, the typical final masses of the stars which form are likely between
$\sim$ 10 and $\sim$ 10$^3$ M$_{\odot}$ (e.g. Hirano et al. 2014).
The outcome of the evolution of a large fraction of these stars
is expected to be the formation of a BH with a mass that is a
substantial fraction of the initial stellar mass (e.g. Fryer et al. 2001; Heger et al. 2003). 

\begin{figure}
\begin{center}
\includegraphics[width=3.0in]{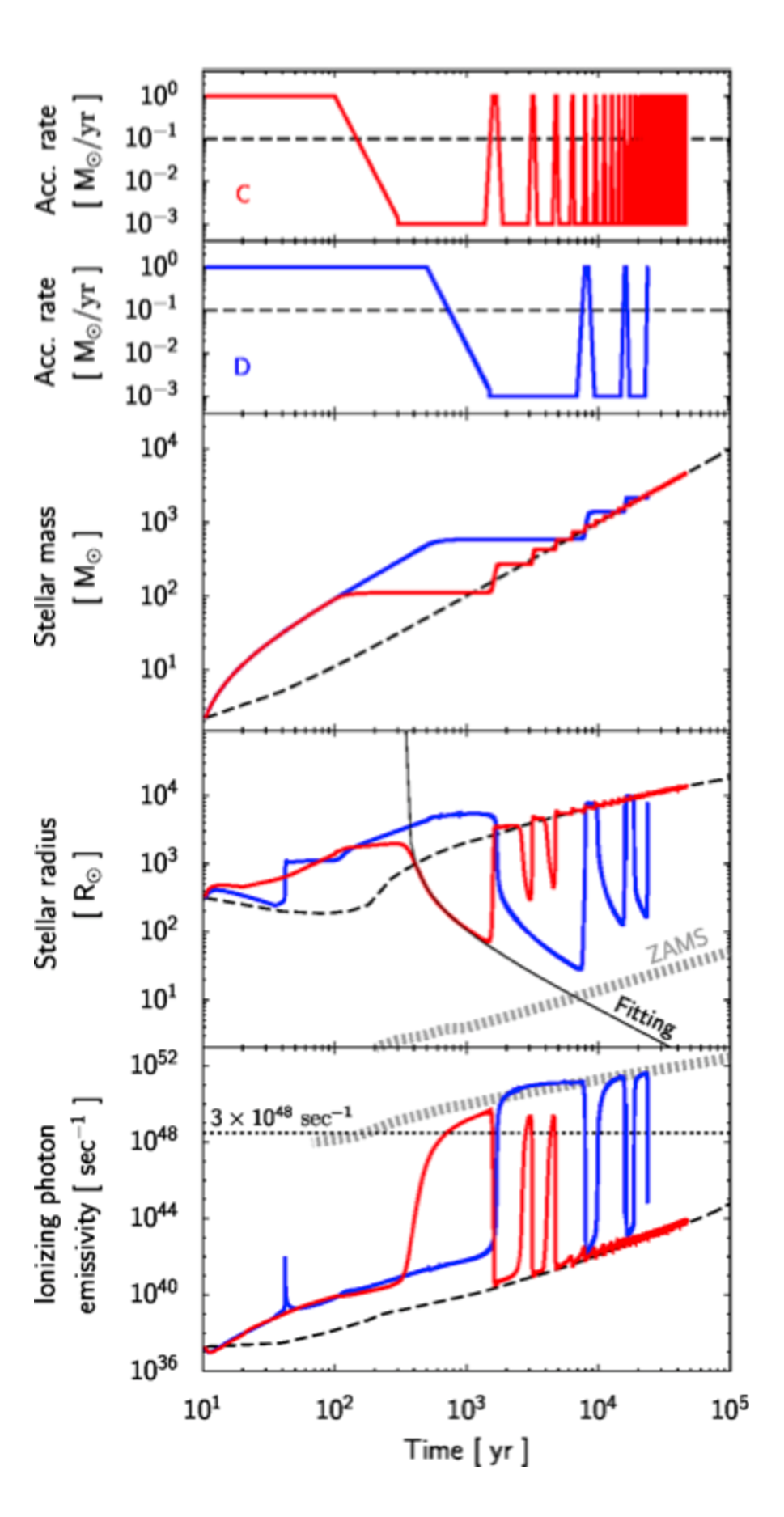}
\caption{The properties of a growing supermassive star, as functions
  of time, for two different assigned time-varying accretion rates
  (shown by red and blue lines).  During periods of slow accretion the
  stellar radius shrinks, the emitting surface becomes hotter, and the
  rate of ionizing photon emission increases.  However, the impact of
  the ionizing radiation on the accretion flow is not sufficient to
  stop the growth of the star.  From Sakurai et al. (2015b).}\label{Fig3}
\end{center}
\end{figure}

If by some means H$_{\rm 2}$ formation is suppressed, then
atomic hydrogen is left as the principle coolant.  In this case, the
gas cools only to $\sim$ 10$^4$ K and is expected to collapse at
extremely high rates of up to $\sim$ 1 M$_{\odot}$ yr$^{-1}$ into a
supermassive star (SMS), as shown in
Figure 1 (e.g. Koushiappas et al. 2004; Lodato \& Natarajan 2006; Wise et al. 2008; Choi et al. 2013, 2015).  The typical final masses of such SMSs are
expected to be $\sim$ 10$^5$ M$_{\odot}$, orders of magnitude larger
than primordial stars formed from gas cooled by molecular hydrogen
(e.g. Latif et al. 2013a).  The typical final outcome of the evolution of such
massive objects is the formation of a BH with a mass which is a
significant fraction of the SMS itself (e.g. Shibata \& Shapiro 2002;
Begelman 2010; Ball et al. 2011; Fryer \& Heger 2011),
although in some cases it may be much lower due to winds that are
generated during the collapse of the star (e.g. Dotan et al. 2011;
Fiacconi \& Rossi 2015).

While it is possible that a variety of processes can lead to the
suppression of gas cooling and, in turn, to the formation of SMSs
(e.g. Sethi et al. 2010; Inayoshi \& Omukai 2012; Tanaka \& Li 2013; Inayoshi et al. 2015a),
the process which is perhaps most likely to play a decisive role is the 
destruction of H$_{\rm 2}$ molecules by intense ultraviolet radiation 
which is absorbed in the Lyman-Werner (LW) bands of the molecule 
(e.g. Haiman 2013; Fernandez et al. 2014; Visbal et al. 2014a; Latif et al. 2014a).   
For H$_{\rm 2}$ cooling to be effectively suppressed, however,
requires extraordinarily intense LW radiation (e.g. Bromm \& Loeb
2003).  This suggests that SMS formation occurs only in relatively rare regions of
the early universe, in pristine dark matter (DM) halos that are nearby
galaxies which harbor stars producing sufficient LW radiation 
(e.g. Dijkstra et al. 2008, 2014; Agarwal et al. 2012, 2014; Visbal et
al. 2014b; Habouzit et al. 2016; but see Petri et
al. 2012).

Estimates of the number density of SMSs, and of the seed BHs into
which they collapse, are difficult to make, due to the sensitivity of
the required level of LW radiation to the state of the collapsing gas
(e.g. Wolcott-Green et al. 2011; Hartwig et al. 2015a) and to
uncertainties in the rates of key chemical reactions (e.g. Glover
2015).  In addition, as H$_{\rm 2}$ molecules typically form from
H$^{-}$ and H$_{\rm 2}$$^+$ ions, there is a strong sensitivity as
well to the shape of the stellar spectrum producing the LW radiation,
as these ions can be destroyed by lower-energy photons (e.g. Shang et
al. 2010; Sugimura et al. 2014, 2015; Agarwal \& Khochfar 2015; Latif
et al. 2015a).  Also, the same galaxies which produce LW radiation
may also be sources of X-rays, which can penetrate deeply into
collapsing halos and ionize the gas, leading to enhanced H$_{\rm 2}$
formation rates (e.g. Machacek et al. 2003; Glover \& Brand 2003) and,
in turn, to suppression of SMS formation (e.g. Inayoshi \& Omukai
2011; Inayoshi \& Tanaka 2015).  Even ionizing radiation, at
energies only slightly higher than LW radiation, can heat and ionize
the gas at early phases in the collapse, leading to further
suppression of SMS formation (e.g. Johnson et al. 2014; Yue et
al. 2014; Regan et al. 2015).

Finally, it is worth noting that, while the impact of radiation on the prevalence
of SMSs is complex, even moderate levels of LW radiation can lead to
enhancements in the masses of primordial stars which collapse into
seed BHs (e.g. O'Shea \& Norman 2007; Xu et al. 2013; Latif et
al. 2014b; Hirano et al. 2015).
Next, we turn to discuss other factors which determine the final
masses of BH progenitors in the early universe.

\subsection{Expected Mass Scale of Black Hole Progenitors}
The expected initial masses of the first seed BHs are determined largely by
the masses to which their progenitor stars grow during their brief
lifetimes.  In turn, a number of processes can dictate how large these
progenitors grow, depending on the environment in which they form.  

The initial mass function of primordial stars formed from gas
cooled by H$_{\rm 2}$ in DM minihalos is likely set at the smallest
mass scales by the ejection of sub-solar protostars by gravitational
interactions, and at the largest scales by radiative feedback from the
stars themselves which eventually halts their accretion of gas (see
e.g. Bromm 2013; Greif 2015).  Recent radiation hydrodynamics simulations
with cosmological initial conditions suggest that the upper end of the
initial mass function of such stars extends up to $\sim$ 10$^3$ M${\odot}$ (Susa et
al. 2014; Hirano et al. 2014, 2015).  Such massive primordial stars
are expected to collapse to seed BHs with similar masses (e.g. Madau
\& Rees 2001; Schneider et al. 2002; Heger et al. 2003; see also Islam et al. 2003).

If the first generations of stars form in tightly-bound clusters it is possible that mergers may drive
the most massive stars to masses in excess of 10$^3$ M$_{\odot}$
(e.g. Portegies Zwart et al. 2004; Devecchi \& Volonteri 2009; Davies et al. 2011; Katz et al. 2015).  This avenue of BH 
formation from slightly metal-enriched star clusters (e.g. Lupi et
al. 2014) can also produce seeds which may subsequently grow to larger,
supermassive scales (e.g. Devecchi et al. 2012; Yajima \& Khochfar 2015). 

The expected mass scale of much more massive SMSs formed from accretion of
atomically-cooled primordial gas is likely set either by their
limited lifetimes (e.g. Begelman 2010; Johnson et al. 2012) or by the
depletion of the gas reservoirs from which they accrete
(e.g. Schleicher et al. 2013).  The radiative
feedback from these objects is expected to be relatively weak, 
owing to the low temperatures of the bloated outer envelopes 
characteristic of such rapidly accreting objects (e.g. Hosokawa et
al. 2012, 2013; Begelman 2012).  Even under intermittent high accretion rates, as
expected due to accretion from a clumpy disk (e.g. Inayoshi \& Haiman
2014; Latif \& Volonteri 2015; Sakurai et al. 2015a; Luo et al. 2015), rapid
gas infall appears to drive the expansion of the star episodically,
limiting the emission of ionizing photons, as shown in Figure 2 
(Sakurai et al. 2015b).  Likewise, high accretion rates are likely
to be realized even in highly turbulent atomically-cooled
collapsing gas (e.g. Prieto et al. 2013; Van Borm et al. 2013, 2014).  
Once the conditions are met for the formation of a SMS, it thus appears
there is little to limit its growth to a characteristic mass scale of
at least $\sim$ 10$^5$.  The fate of most such massive objects is to collapse to a
BH with a similarly large mass, although a large fraction of
its mass may instead be lost if sufficiently large accretion rates
persist up until its collapse (e.g. Begelman et al. 2006; Dotan et
al. 2011; Schleicher et al. 2013). 

An avenue for the formation of even more massive seed BHs, with
initial masses up to 10$^8$ - 10$^9$ M$_{\odot}$, has been
suggested to be rapid gas inflow into the center of rare, Milky Way-sized
galaxies triggered by gas-rich mergers at high redshift (Mayer et al. 2010; Bonoli et
al. 2014; Mayer et al. 2015).  While the nature of the initial
conditions in this scenario is more complex than those based on
the collapse of primordial gas in less massive DM halos, and may
involve complications such as the pre-existence of BHs in the centers
of the merging galaxies, somewhat idealized simulations suggest that
heavy, optically thick accretion flows may well lead to the formation of
such massive central objects (Mayer et al. 2015; but see also Ferrara et al. 2013).

\section{Initial Growth of Seed Black Holes}
The rate at which seed BHs grow, immediately following their
formation, is dictated both by the mechanical and radiative feedback from
their progenitor stars and by the impact of the radiation produced in
the accretion process itself.  Here we discuss these processes and the
basic expectations for the early growth of BH seeds.  

\subsection{Impact of Radiation Emitted from the Progenitor}
BH progenitors being massive stars with large luminosities, in many
cases the radiation they emit dramatically alters their surroundings.
In turn, as it is from these surroundings that the nascent BHs
accrete, the radiation from progenitor stars can play a key role in
determining how quickly the first BHs grow.  

\begin{figure}
\begin{center}
\includegraphics[width=0.9\columnwidth]{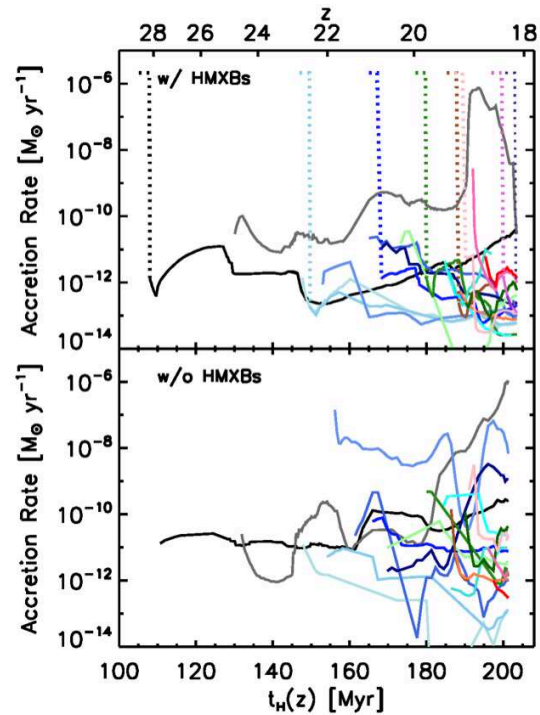}
%\includepdf{accrete.pdf}
\caption{Accretion rates of BHs in simulations with (top) and without
  (below) radiative feedback from high mass X-ray binaries.  In
  conjunction with the radiative feedback from their stellar
  progenitors, the impact of this X-ray feedback is to dramatically
  limit the accretion of gas onto BHs formed from primordial stars
  formed from H$_{\rm 2}$-cooled gas in minihalos. From  Jeon et al. (2014).}\label{Fig1}
\end{center}
\end{figure}

Massive primordial stars formed from H$_{\rm 2}$-cooled gas in
minihalos are expected to produce copious amounts of ionizing radiation for
at least 1 - 2 Myr before their collapse (e.g. Schaerer 2002).  The gas
surrouding them is therefore heated to well above the virial
temperature of their host DM halos, and a large, low-density H~II
region is left when the star collapses  (e.g. Kitayama
et al. 2004; Whalen et al. 2004; Alvarez et al. 2006; Abel et al. 2007).  
Such BHs are thus expected to be born into very low-density
environments, from which gas accretion is exceedingly slow, leading to
a significant delay of perhaps up to $\sim$ 100 Myr in the growth of
these seeds (Johnson \& Bromm 2007; Alvarez et al. 2009; Johnson et
al. 2013a).  

In contrast, the radiative feedback from SMSs is likely to be much
weaker, due to their having cool surfaces from which little ionizing
radiation is emitted (e.g. Hosokawa et al. 2013; Schleicher et
al. 2013; Sakurai et al. 2015a,b).  This is likely to result in a
large reservoir of dense gas from which such a so-called direct collapse
black hole (DCBH) can accrete.  Cosmological simulations of this process suggest that
Eddington-limited accretion is possible starting from the formation of
the DCBH (Johnson et al. 2011; Aykutalp et al. 2014), and more idealized models of the
accretion flow at smaller scales near the DCBH suggest that even
higher rates of growth are attainable (e.g. Pacucci \& Ferrara 2015;
Inayoshi et al. 2015b).   Thus, not only do SMSs produce large initial
BH seeds, these seeds are also likely to grow rapidly starting from 
birth.

\begin{figure}
\begin{center}
\includegraphics[width=0.75\columnwidth,angle=90]{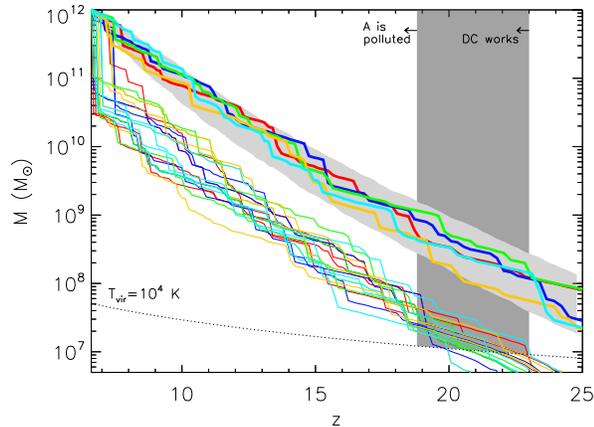}
\caption{The distribution of possible masses of the DM halo hosting
  the candidate DCBH in the CR7 system (thin lines) and the
  corresponding masses of the larger DM halo with which it merges, as
  functions of redshift $z$.  The mass of a DM halo with a virial
  temperature of 10$^4$ K is denoted by the dotted line.  The LW
  radiation emitted from stars formed in the larger halo is thought to
  destroy the H$_{\rm 2}$ molecules in the smaller one until its virial temperature exceeds 10$^4$ K, at which point the gas collapses rapidly and forms a supermassive star that later collapses into a DCBH.  From Agarwal et al. (2015).}\label{Fig3}
\end{center}
\end{figure}

\subsection{Impact of Progenitor Explosion}
The supernovae (SNe) which mark the end of life of many massive
primordial stars dramatically impact their surroundings, blowing out
the gas from the centers of their host DM halos, thereby limiting
the gas reservoir available for accretion onto any seed BHs that are
left behind.

Primordial stars with masses of 140 - 260 M$_{\odot}$ are expected to
explode as pair-instability SNe, with energies of $\sim$ 10$^{53}$ erg
(e.g. Heger et al. 2003).  Such energetic explosions can easily
evacuate the gas from the centers of DM minihalos (e.g. Greif et
al. 2007; Vasiliev et al. 2008; Wise \& Abel 2008).  While such SNe
leave behind no BH remnant, it is likely that primordial stars in this
environment form in clusters (e.g. Greif et al. 2015) and such strong
mechanical feedback can leave the BH remnants of other stars to
accrete from low-density gas, resulting in very slow growth.
Lower-mass primordial stars may also explode as less energetic SNe,
which may or may not effectively evacuate their host minihalos of gas
for long periods, depending on the energy of the explosion and on the
mass of the minihalo (Kitayama et al. 2005; Whalen et al. 2008; and
Ritter et al. 2012).  Indeed, it is expected that a large fraction of
primordial SNe may leave behind a remnant BH, which may accrete gas
that is not evacuated in the explosion, although some fraction of
these BHs may be relegated to the low-density IGM due to natal
kicks imparted to them in assymetric SNe (Whalen \& Fryer 2012).

It is also a possibility that some SMSs explode in extremely energetic
SNe (e.g. Montero et al. 2012; Chen
et al. 2014).  While these violent events may not leave behind a
remnant BH, it is possible that some SMSs form in binary systems
(Regan \& Haehnelt 2009; Reisswig et al. 2013; Whalen et al. 2013c; Latif et
al. 2015b; Becerra et al. 2015) and
accretion onto a BH formed from the collapse of a companion SMS would
likely be strongly inhibited due to the evacuation of high density
gas from the center of the host halo (Johnson et al. 2013b; Whalen et
al. 2013a,b).  Finally, if a relativistic jet is launched from the
center of a collapsing SMS, perhaps powering a gamma-ray burst, it
could also drive out the surrounding gas, resulting in a similar
suppression of accretion onto the DCBH that is left behind (Matsumoto et al. 2015).

\subsection{Impact of Radiation Emitted in the Accretion Process}
The ineraction between the accretion flow onto a BH and the radiation
emitted in the accretion process is highly complex, and is
sensitive to the initial conditions of the flow as well as to physical
effects on both small scales near the BH and larger scales where the
flow originates.    

In cases where a seed BH is accreting from a relatively low-density
medium, such as occurs when the radiative and/or mechanical feedback
from the progenitor effectively clears out the gas, the high-energy
radiation that is generated in the accretion disk can travel large
distances (e.g. Kuhlen \& Madau 2005), heating and further expelling the gas.
This can result in a dramatic suppression of accretion onto the BH seeds
left by primordial stars formed via H$_{\rm 2}$ cooling in DM minihalos (e.g. Alvarez et
al. 2009), and can be exacerbated if there is additional radiative
feedback from X-ray binaries (e.g. Mirabel et al. 2011; Fragos et
al. 2013;  Jeon et al. 2014; Xu et al. 2014; Chen et al. 2015; Ryu et
al. 2015), as shown in Figure 3.  

\begin{figure}
\begin{center}
\includegraphics[width=\columnwidth]{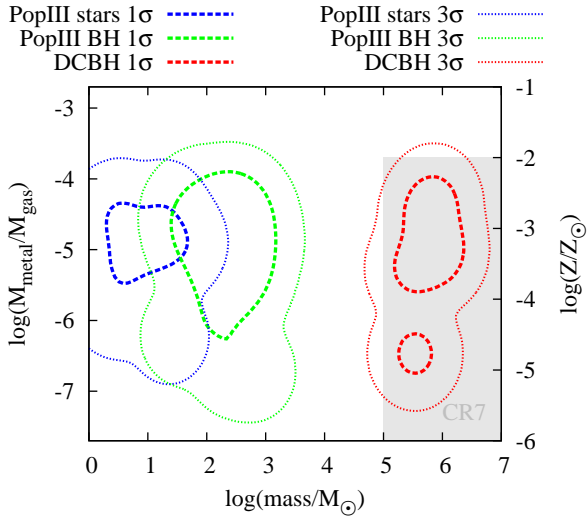}
\caption{The metallicity of the bright Lyman-$\alpha$ emitter in CR7, as
  produced by modeling its emission as due to a 
  cluster of primordial stars (blue), an accreting BH formed from the collapse of a
  primordial star in a minihalo (green), and an accreting
  DCBH (red), with the masses of these respective objects shown along
  the horizontal axis.  The thick and thin dashed contours correspond to
  the range of properties found in 68  and 99 percent, respectively, of the model realizations studied.  The gray region is that in which the properties are consistent with CR7; only an accreting DCBH can simultaneously produce the bright emission observed while also suppressing metal enrichment strongly enough so as to limit the strength of metal emission lines to a level consistent with the observations.  From Hartwig et al. (2015b).}\label{Fig3}
\end{center}
\end{figure}

In cases where a seed BH is accreting from relatively dense gas, as is
expected to occur for DCBHs born in dense regions little affected by
feedback from the progenitor SMS,
growth via accretion is much more rapid.  Cosmological simulations 
resolving the Bondi radius of DCBHs and including ionizing radiation 
(Johnson et al. 2011) and X-ray (Aykutalp et al. 2014) feedback,
suggest that the accretion rate may at times peak near the
Eddington rate, although it is on average well below this value.  Similar
periodic behavior is found in more idealized calculations which resolve smaller
scales near the BH, as well (e.g. Milosavljevi{\' c} et al. 2009a,b;
Park \& Ricotti 2011, 2012).  While this suggests that accretion onto
DCBHs may occur at, on average, somewhat sub-Eddington rates, if
accretion proceeds from sufficiently dense gas then radiation from the
accretion disk can be stymied and the accretion rate may exceed the
Eddington rate for long periods (e.g. Inayoshi et al. 2015; Pacucci et al. 2015).
Given that there is a strong dependence of the final BH mass on the early
accretion rate, it is critical to determine the state of the
environment in which a DCBH is formed in order to model its subsequent
evolution and fate.

Intriguingly, the recently detected bright Lyman-$\alpha$ source in
the CR7 system (Sobral et al. 2015) may provide strong empirical
constraint on the feedback resulting from accretion onto DCBHs.
Modeling of this source, in which no emission lines from heavy
elements have been detected, suggests that it is likely to be powered
by a DCBH accreting primoridial gas (e.g. Pallottini et al. 2015).
This is consistent with an observed nearby galaxy which may have produced the
LW radiation required for DCBH formation (Agarwal et al. 2015; see
also Regan et al. 2015), as shown in Figure 4.  This is also
consistent with modeling that suggests that lower-mass seed BHs formed
in minihalos would not likely evolve into a system consistent with the data
(Hartwig et al. 2015b), as shown in Figure 5.  The lack of heavy
elements inferred for this source implies that star formation has been
suppressed continually for at least $\sim$ 10$^8$ yr (Pallottini et al. 2015)
and perhaps longer (Agarwal et al. 2015).  This may be explained by suppression of H$_{\rm 2}$
formation due to the intense LW radiation emitted from a DCBH
accreting continually at a substantial fraction of the Eddington rate
(e.g. Johnson et al. 2011) in conjunction with the strong LW radiation
emitted from its neighboring galaxies, although a countervailing
effect is that the X-rays emitted from the accretion disk may promote 
molecule formation and star formation (Aykutalp et al. 2014), as shown in Figure 6.
Cosmological simulations directed at modeling the formation of systems specficially like
CR7 will help to shed more light on its nature.

\begin{figure*}
\begin{center}
\includegraphics[width=4.2in,angle=90]{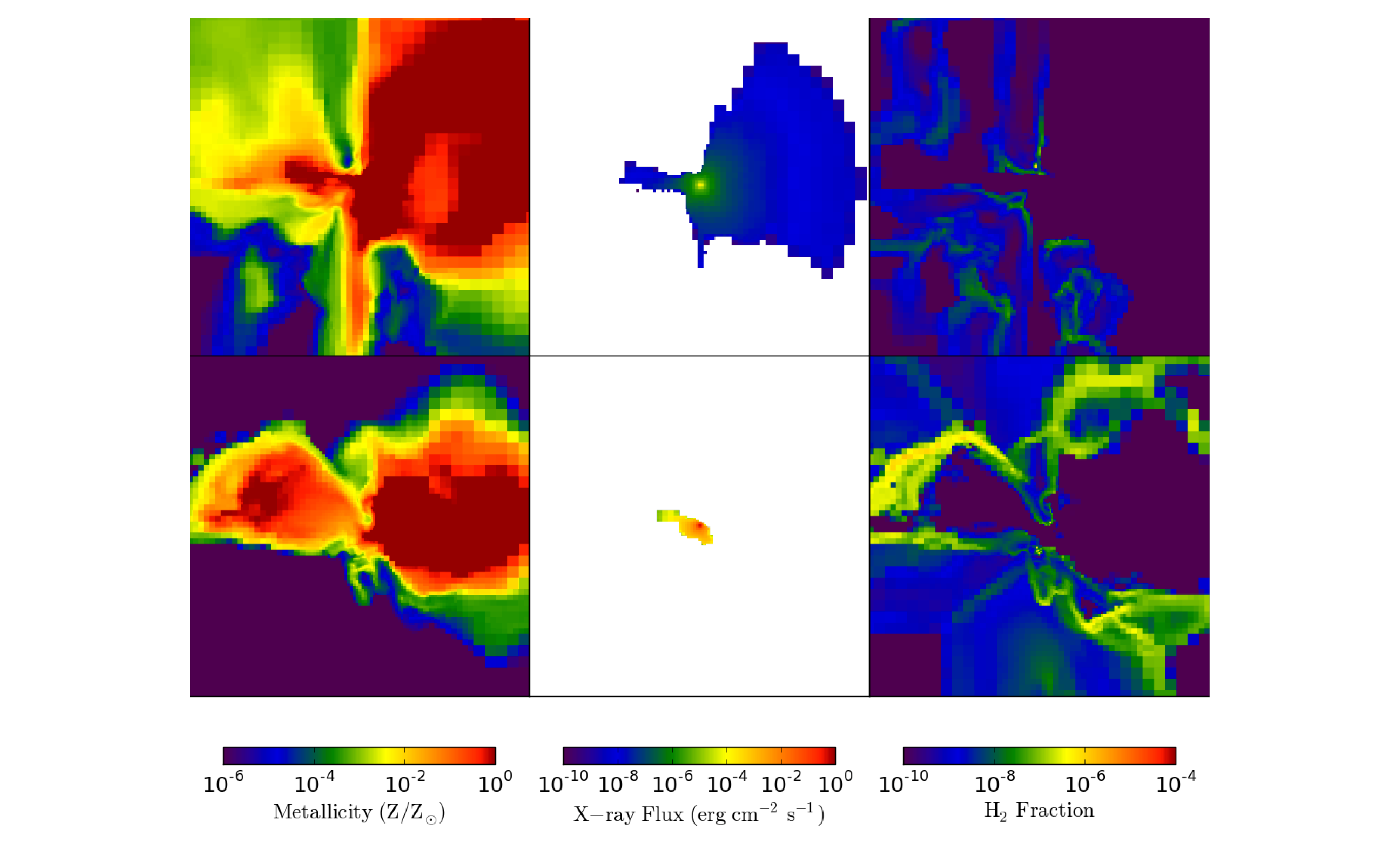}
\caption{The metallicity (left), the X-ray flux (middle), and the
  H$_{\rm 2}$ fraction (right) within a 1 kpc region surrounding a
  central 5 x 10$^4$ M$_{\odot}$ accreting DCBH.  With a background LW
  radiation field (with $J_{\rm 21}$ in units of 10$^{-21}$ erg s$^{-1}$ cm$^{-2}$ Hz$^{-1}$ sr$^{-1}$) of $J_{\rm 21}$ = 10$^3$ (bottom panels) the gas is
  able to cool and collapse more readily than with $J_{\rm 21}$ =
  10$^5$ (top panels), due to stronger cooling by H$_{\rm 2}$
  molecules.  As a result, the BH accretes more rapidly in the former case, although the X-rays it emits do not propagate as widely as in the latter.  From Aykutalp et al. (2014).}\label{Fig3}
\end{center}
\end{figure*}

\section{Super-Eddington Accretion as a Path to Supermassive Black Holes}
While the initial mass of a BH seed has a large impact on the final
mass to which it can grow via accretion at or below the Eddington
limit, if accretion proceeds from a sufficiently dense reservoir of
gas this strong dependence on the initial mass can be eased.
This possibility was
discussed, among others, by Volonteri \& Rees (2005), who envisaged an
early phase of super-critical accretion in the growth of
SMBHs. Within primordial halos with virial temperature $\gsim 10^4$K,
in particular, conditions exist for the formation of a rotationally supported disk at the center of the halo. The authors showed that if the disk rotates as a solid body, then a tiny accretion disk can form inside the radiation trapping radius, and hence the BH can accrete most of the infalling gas.  On similar theoretical grounds, Kawakatu \& Wada (2009) proposed a simple model for the co-evolution of a clumpy  circumnuclear disk and of a SMBH, and showed that the observed bright QSOs at $z\gsim 6$ are best explained by a late growth ($z\lsim 10$) of the BH seed at super-Eddington (SE) rates. This is consistent with Wyithe \& Loeb (2012), who found that a large fraction of galaxies at $z\gsim 6$ may exhibit Bondi accretion rates large enough to trap the produced photons, hence allowing for SE accretion. The authors pointed out that photon trapping can operate only for BHs with masses 
$\lsim 10^5 M_\odot$, as for larger holes the Bondi accretion radius exceeds the disk scale height. 

\begin{figure*}
\begin{center}
\includegraphics[width=3.0in,angle=-90]{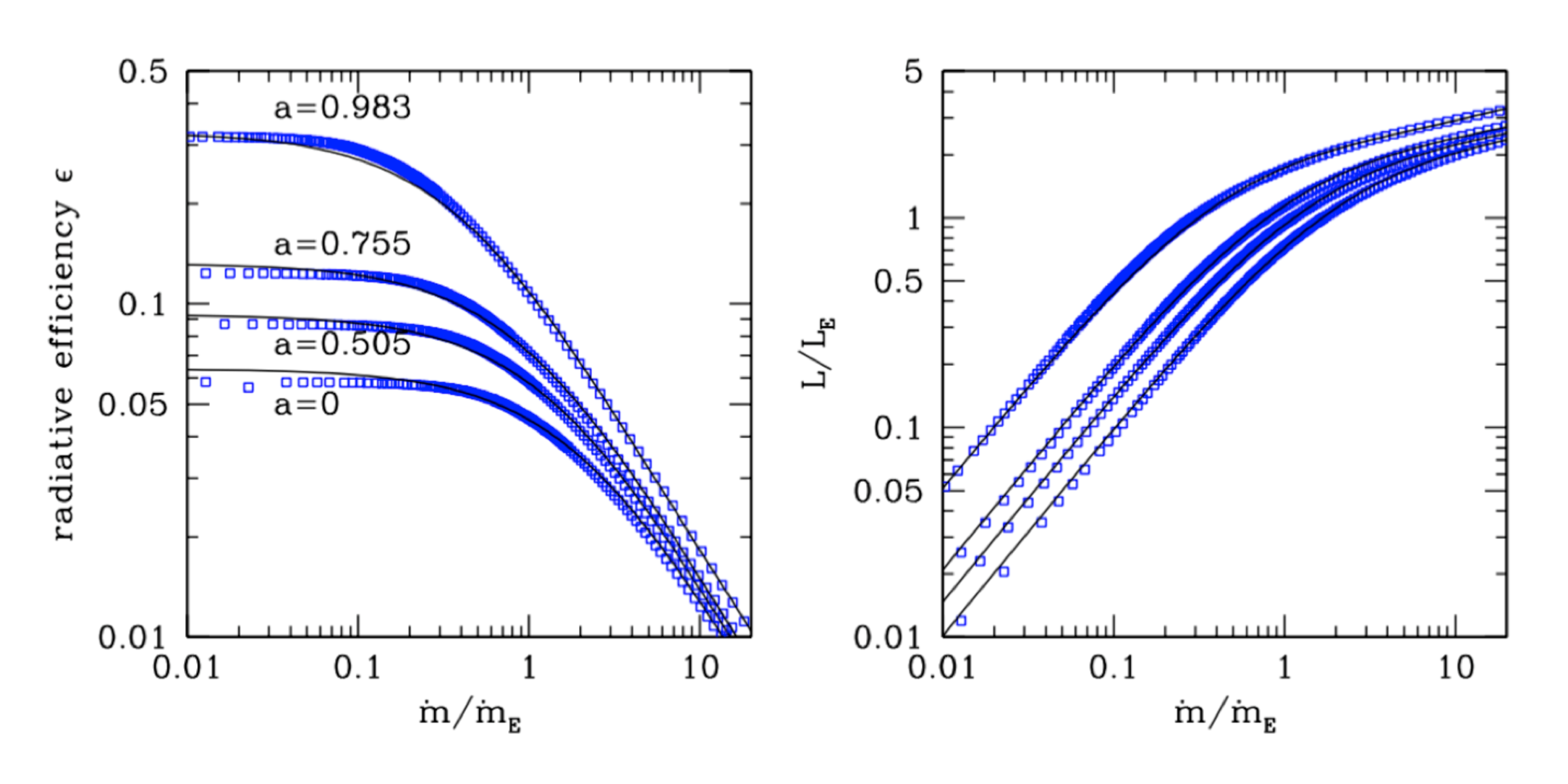}
\caption{Radiative efficiency and total luminosity  of an accreting
  BH are plotted in the left and right panels, respectively, as a function of
  the accretion rate $\dot M$ (in units of the Eddington rate). The blue points are the results of the numerical integration of the relativistic slim disk equations obtained by Sadowski (2009), while the solid curves from top to bottom show  best-fit functions. From Madau et al. (2014).}
\label{Fig:SE1}
\end{center}
\end{figure*}

Observationally, evidence for near-Eddington or SE flows has been
accumulating in recent years.  A study of a large sample of active
galactic nuclei (AGNs) suggests that many of them emit considerably
more energy and have higher Eddington ratios than previously assumed
(Netzer \& Trakhtenbrot 2014). Kormendy \& Ho (2013) have argued that
the normalization of the local BH scaling relations should be
increased by a factor of five to $M_{\rm BH} = 0.5\%$ of the bulge
mass. This increases the local mass density in BHs by the same factor,
decreases the required mean radiative efficiency to 1 - 2\%, and may
be evidence for radiatively inefficient SE accretion (e.g., Soltan
1982; Novak 2013; see also Li 2012). At high redshifts, the very soft
X-ray spectrum of ULAS J1120+0641 appears to suggest that this quasar
is accreting at super-critical rates (Page et al. 2013). Moreover,
super-critical accretion onto stellar-mass BHs has been invoked to
explain the nature of the ultraluminous X-ray sources (e.g., Gladstone
et al. 2009; Middleton et al. 2013).

The properties of flows accreting onto BHs at SE rates were first investigated by Abramowitz et al. (1988), who found the so-called "slim-disk solution" to the Navier-Stokes equations. The standard Shakura-Sunyaev disc solution (Shakura \& Sunayev 1973) assumes that the viscosity-generated heat is locally radiated away as a black body. It is effectively a zero-dimensional model, described by algebraic equations valid at any particular radial location in the disk, independent of the flow 
physical conditions at larger or shorter radii. As already pointed out in the original work by Shakura \& Sunayev, 
at high accretion rates, i.e., when the luminosity produced is larger than $\simeq 30\%$ the Eddington limit, the local 
assumption is no longer valid. Abramowitz et al. (1988) derived a more general set of 
non-local solutions valid in the high accretion rate regime. The slim disk  
(sometimes nicknamed "the Polish doughnut") has the form of a
stationary, optically and geometrically thick accreting flow, and it
is obtained by numerical integration of the two-dimensional stationary
Navier-Stokes equations with a transonic point. Such a solution allows for the generated 
heat to be advected in by the accreting gas. In simple terms, in slim disks the flow opacity is so high that the diffusion timescale of photons exceeds the accretion time scale. Part of the radiation produced is then ``trapped" within the accreting flow, and fated to feed  the BH (i.e., to add to its mass). The reduced radiative efficiency opens the possibility of having extremely high
mass accretion rates unimpeded by radiation, well above the Eddington limit. 

Among the most advanced studies to date of the Polish doughnut are those by Sadowski and collaborators (see, e.g., Sadowski 2009, Sadoswki et al. 2014), who performed MHD general relativistic axisymmetric simulations of SE accretion flows. 
State-of-the-art 3-D simulations with full radiative transfer (Jiang
et al. 2014; McKinney et al. 2014) show how vertical advection of radiation caused by
magnetic buoyancy transports energy faster than photon diffusion,
allowing a significant fraction of the photons to escape from the
system before being advected into the hole. In the context of early BH seed
growth, the interesting properties are: i) even for the somewhat
larger (compared to, e.g., Sadowski's studies) radiative efficiencies
found by Jiang et al. (2014), radiation does not halt accretion, as the
outflowing gas and photons leave the system through a  funnel near the
rotation axis (while mass is accreted in the plane of the disk); and
ii) the radiative efficiency at SE rates is almost independent of
the BH spin. As highlighted by Volonteri et al. (2015), this has the
important consequence that the expected BH spin-up caused by the
accreted gas would not hinder the BH growth through a high radiative
efficiency, as shown in Figure 7.

\begin{figure*}
\begin{center}
\includegraphics[width=4.0in,angle=-90]{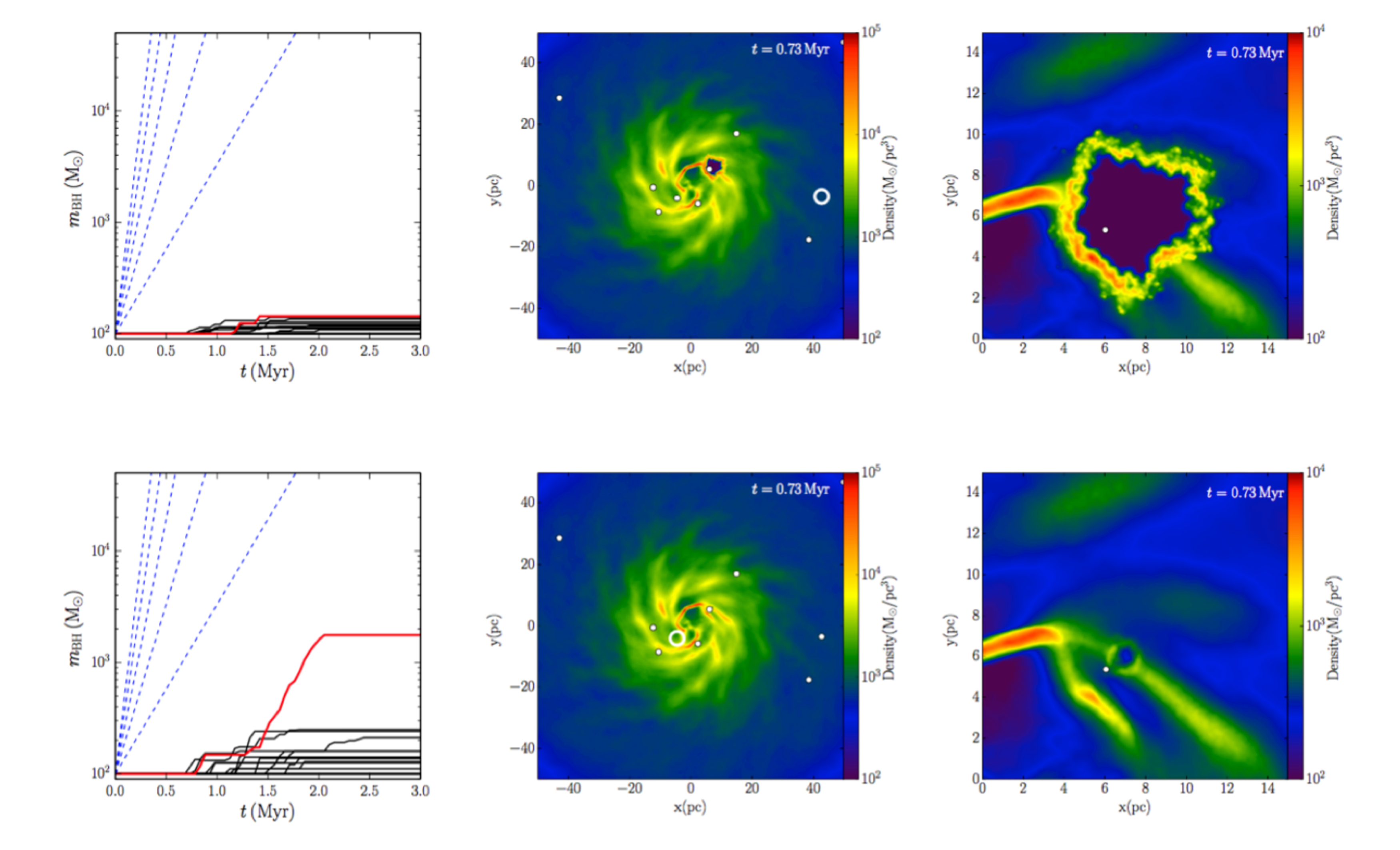}
\caption{Left panels: BH masses as functions of time, assuming 10\% radiative efficiency of accretion (top), and the radiatively inefficient slim-disk solution (bottom). The red lines correspond to the most massive BHs at the end of the runs, while the blue dashed lines trace accretion histories at fixed Eddington ratios of 500, 400, 300, 200 and 100, respectively. Central panels: gas density maps for the two runs at t = 0.73 Myr. Right panels: zoom in of a region heated by BH feedback. The white dots mark the positions of the BHs. From Lupi et al. (2015).}
\label{Fig:SE2}
\end{center}
\end{figure*}

SE accretion of stellar-sized BHs has been recently recognized as a
possible alternative to explain the existence of billion
solar mass BHs at redshifts as high as $\simeq 7$ (Madau et
al. 2014; Tal \& Natarajan 2014; Volonteri et al. 2015). Volonteri et al. (2015) made specific predictions that long-lived SE accretion occurs only in galaxies with copious low-angular momentum gas, and pointed out how the duty-cycle of a BH can be as low as $\sim 0.01$, to be compared to $\sim$ unity required if accretion proceeds at 
a sub-critical pace. Madau et al. (2014) used the numerical solutions
provided by Sadowski (2009) in terms of radiative efficiency
vs. accretion rate, and showed that, under the assumption that a gas
supply rate a few times the Eddington rate ($\gsim$ $0.01 M_5$
M$\odot$ yr$^{-1}$ (where $M_5$ is the mass of the hole in units of
$10^5$ $M_\odot$) is indeed able to reach the central MBH for a period
of 20 Myr or so,  a few episodes of SE accretion may turn a light seed
hole into a rare bright quasar, such as those observed at $z\sim 6$ by
the SDSS. The high rates required may be determined both by local physics within the accretion flow (e.g., Dotan \& Shaviv 2011) as well as by the large-scale cosmological environment in which MBHs and their hosts are growing (e.g., Mayer et al. 2010). 
By means of a semi-analytical approach, Smole et al. (2015) found that low-mass seeds ($M\simeq 100 M_\odot$) accreting at 3-4 times the Eddington rates could in principle explain the mass function of SMBHs at $z\sim 7$. 

The key aspect to assess the importance of SE accretion in the early
growth of BHs is the mass supply from large (kpc) scales, and the ability of stellar-size BHs to intercept and accrete such inflowing gas. Two approaches have been pursued so far. Small scale radiation hydrodynamical simulations, and larger scale hydro simulations in a cosmological framework.

Recently, Pacucci et al. (2015) and Inayoshi et al. (2015) elaborated
upon this concept through 1-D and 2-D radiation hydro-dynamical
simulations of a converging flow, and found that, in certain conditions, rapid gas supply at SE rates 
from the Bondi radius can occur without being impeded by radiative
feedback. In this regime accretion is steady, and larger than 3,000
times the Eddingtion rate (defined there as $\dot m_{\rm Edd}\equiv
L_{\rm edd}/c^2$). As originally found by Milosavljevi{\' c} et al. (2009)
and by Park \& Ricotti (2011), at lower Bondi rates accretion is
instead episodic, because of radiative feedback. A 2-D analysis,
performed by Yang et al. (2014), showed the additional importance of radiation in stabilising convection in slim disks. 

Though extremely detailed, such simulations do not capture the
dynamics of the gas feeding the BH from large scales. As an example, 
the 3-D simulations by Jiang et al. (2014) cover only a small region well within the trapping radius. 
For such reasons, larger scale simulations of the 3-D circum-nuclear
disk have been performed. The drawback of this approach to the problem
is that simulations on $\gsim$ 100 ${\rm pc}$ scale can not resolve the small accretion disk around the BH, and hence one must employ  
"subgrid recipes" to model the gas down to the accretion radius. 
Given that feedback from primordial stars and their BH remnants tends
to strongly limit accretion (e.g. Alvarez et al. 2009), stellar-size BH seeds may not be prone to early growth (and incidentally, may not be shining as miniquasars). 

The situation is likely to be quite different in more evolved galaxies, where gas is thought to form a circum-nuclear disk. 
In this context, very recently Lupi et al. (2015) studied the selective accretion of stellar-mass seeds 
in the gaseous circum-nuclear disks expected to exist in the cores of
high redshift galaxies. Their sub-pc resolution hydrodynamical
simulations showed that stellar-mass holes orbiting within the central
100 pc of the circum-nuclear disk bind to very high density gas clumps
that arise from the fragmentation of the surrounding gas. Owing to the
large reservoir of dense cold gas available, a stellar-mass BH growing
at SE rates according to the Òslim discÓ solution can increase its
mass by 3 orders of magnitudes within a few million years. The low
radiative efficiency of super-critical accretion flows is instrumental
to this rapid growth, as they imply modest radiative
heating of the surrounding nuclear environment.  The left panels of
Figure 8
show BH masses as functions of time, while central and right panels compare the case of SE accretion with a standard case of 10\% radiative efficiency. 
The results of Lupi et al. (2015) indicate that a short phase of SE accretion onto
stellar-size BHs in the cores of star-forming galaxies at $z\gsim 10$
may be a first step toward the formation of the supermassive BHs powering the brightest QSOs observed at $z\lsim 7$.

\begin{figure}
\begin{center}
\includegraphics[width=\columnwidth]{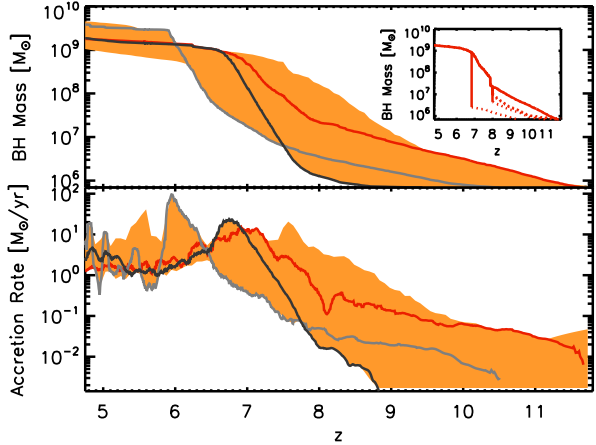}
\caption{The accretion rate (bottom) and mass (top) of three BHs, as
  functions of redshift $z$, as found in a large-scale cosmological
  simulation.  These BHs are seeded with masses consistent with
    DCBHs, grow to masses $\ge$ 10$^9$ M$_{\odot}$ before $z$ = 6, and are broadly consistent with observations of quasars at these redshifts.  From Di Matteo et al. (2012).}\label{Fig3}
\end{center}
\end{figure}

\section{Implications for the Fate of the First Massive Black Holes}
In order to explain the formation of the largest SMBHs inferred to
power quasars at high redshift, it is not only necessary to specify
their initial masses and early accretion rates; their growth over
10$^8$ - 10$^9$ years of cosmic time must be evaluated.
 As the relatively small DM halos in which the first
seed BHs
form merge with many other halos and end up in much more massive
systems, large-scale simulations are required to track their long term
evolution.  In turn, given finite computational resources, such
large-scale simulations generally can not capture accretion of gas and the
associated radiative feedback in great detail (e.g. Curtis \& Sijacki
2015; Rosas-Guevara et al. 2015; but see also Pelupessy et al. 2007) and sub-grid
prescriptions must be relied upon (e.g. Debuhr et
al. 2010; Power et al. 2011; Hobbs et al. 2012).  

Nonetheless, there is some encouraging consistency between models of BH
seed formation and the observational data on high redshift quasars, which is
strengthened by large-scale cosmological simulations of the evolution
linking the former to the latter.  In particular, the masses of BH
seeds which are typically adopted in large-scale
cosmological simulations (e.g. Bellovary et al. 2011; Kim et al. 2011) are
often consistent with the masses expected for seeds formed from the collapse 
of SMSs in the direct collapse model (e.g. Ferrara et al. 2014).
These simulations generally show that the growth of $\sim$ 10$^9$
M$_{\odot}$ SMBHs from these seeds, fueled by accretion of low angular
momentum gas (e.g. Hopkins \& Quataert 2010) delivered by cold flows
(e.g. Dubois et al. 2012, 2013),
could proceed fast enough to explain the highest redshift quasars (Li et
al. 2007; Sijacki et al. 2009, 2015; Di Matteo et al. 2012), as shown
in Figure 9.  

Observations of BHs inhabiting more local, dwarf galaxies (e.g. Reines et al. 2011;
2014; Greene et al. 2012; Mezcua et al. 2015) and Milky Way satellites
(see e.g. van Wassenhove et al. 2010) may also be explained by
the DCBH seed model, although it is likely that other seeding
processes populate such galaxies as well.  Future observations
which constrain the properties of BHs in dwarf galaxies, of the BHs
powering high redshift quasars, and of those which may reside in the
brightest primordial systems, for which CR7 is a strong candidate,
will be critical to determining how the first massive BHs are
seeded and grow in the early universe.

\begin{acknowledgements}
The authors would like to thank Rosa Valiante, Rafaella Schneider and
Marta Volonteri for the opportunity to contribute this review.  Work at LANL was done under the auspices of the National Nuclear Security Administration of the US Department of Energy at Los Alamos National 
Laboratory under Contract No. DE-AC52-06NA25396.
\end{acknowledgements}

\end{document}